\documentclass[sigconf,9pt]{acmart}

\usepackage{amsthm,amsmath,amssymb,array,color,colortbl,graphicx,multirow}

\usepackage{microtype}
\usepackage{comment}
\usepackage{algorithm}
\usepackage{balance}
\usepackage{xspace}
\usepackage{wrapfig}
\usepackage{algorithm}
\usepackage[noend]{algorithmic}
\usepackage{enumitem}

\newtheorem{definition}{Definition}

\newcommand{\Off}{\textsc{Off}}
\newcommand{\On}{\textsc{On}}
\newcommand{\Fixed}{\textsc{Stat}}
\newcommand{\Obl}{\textsc{Obl}}
\newcommand{\Gen}{\textsc{Gen}}

\newcommand{\netw}{N}
\newcommand{\netws}{\mathcal{N}}

\newcommand{\demg}{G}

\newcommand{\insrt}{\mathrm{insert}}

\newcommand{\adjust}{\mathrm{adjust}}

\newcommand{\search}{\mathrm{search}}
\newcommand{\delete}{\mathrm{delete}}
\newcommand{\route}{\mathrm{route}}

\newcommand{\Cost}{\mathrm{Cost}}
\newcommand{\RecCost}{\mathrm{adj}}
\newcommand{\RouCost}{\mathrm{srv}}
\newcommand{\params}{\mathrm{par}}

\def\A{\mathcal{A}}

\newcommand\stefan[1]{\color{blue}\textbf{Stefan: #1 }\color{black}}

\renewcommand\footnotetextcopyrightpermission[1]{}
\setcopyright{none}

\acmDOI{}

\acmISBN{}

\acmPrice{}

\begin{document}

\title{ReNets: The Art of Reconfigurable Network Design}

\title{Self-Adjusting Networks}

\title{ReNets: Reconfigurable Routable Networks}

\title{DDANs: 
Dynamic Demand-Aware Networks}

\title[Dynamic Demand-Aware Networks]{
Routing Meets Entropy Bounds:
\\
Dynamic Demand-Aware Networks}

\title{Theoretical Foundations for Self-Adjusting Networks (SANs)}

\title[Theory of Self-Adjusting Networks]{Toward Demand-Aware Networking: \\A Theory for Self-Adjusting Networks}

\author[Avin and Schmid]{Chen Avin$^1$ \quad Stefan Schmid$^2$\\
\small~$^1$ Ben Gurion University, Israel \quad
\small~$^2$ University of Vienna, Austria
}




\begin{abstract}
The physical topology is emerging as the next frontier
in an ongoing effort to render communication networks
more flexible.
While first empirical results indicate that these flexibilities 
can be exploited to reconfigure and optimize the network toward
the workload it serves and, e.g., providing the same
bandwidth at lower infrastructure cost, 
 only little is known today about the 
 fundamental algorithmic problems underlying the design of reconfigurable networks.
This paper initiates the study of the theory of demand-aware, self-adjusting networks.
Our main position is that self-adjusting networks should be 
seen through the lense of self-adjusting datastructures. Accordingly, 
we present a taxonomy classifying the different
algorithmic models of 
demand-oblivious, fixed demand-aware,
and reconfigurable demand-aware networks, 
introduce a formal model, and identify
objectives and evaluation metrics. 
We also demonstrate, by examples,
the inherent advantage of demand-aware
networks over state-of-the-art demand-oblivious, fixed networks (such as expanders).
\end{abstract}

\maketitle

%
%
%

\section{Introduction}\label{sec:intro}

Data-centric applications, including
online services like web search, social networks, storage, financial services,
multimedia, etc.~\cite{survey2017datacenter},
as well as emerging applications such as distributed machine learning, 
generate a significant amount of
network traffic~\cite{talk-about,kraken,dist-ml,singh2015jupiter,cisco-pro}.
It is hence not surprising that the design of 
efficient datacenter networks has received much
attention over the last years.
The topologies underlying modern datacenter networks 
range from trees~\cite{clos,fat-free} over hypercubes~\cite{bcube,mdcube}
to expander networks~\cite{xpander} and provide high connectivity at low cost~\cite{survey2017datacenter}.

Until now, these networks also have in common that their topology
is \emph{fixed} and \textbf{\emph{oblivious}} to the actual demand (i.e., workload
or communication pattern)
they currently serve.
As such, topologies are designed 
to provide worst-case guarantees,
such as full bisection bandwidth,
and to support arbitrary (\emph{all-to-all}) 
communication patterns~\cite{clos}.

Emerging technologies in like optical circuit switches~\cite{helios,rotornet,reactor,mordia},
60 GHz wireless communication~\cite{zhou2012mirror,kandula2009flyways} and
free-space optics~\cite{firefly,projector} herald a very different kind
of network topologies: malleable topologies which can be quickly 
\emph{reconfigured}~\cite{singla2010proteus}. 
Reconfigurable networks introduce an additional degree of freedom
to the datacenter network design problem~\cite{singla2010proteus,projector,Jia2017,reactor, helios, firefly, zhou2012mirror,augmenting,chen2014osa}.

The technology enables \textbf{\emph{demand-aware networks}} which are optimized toward 
the workload they serve, statically (\emph{fixed} topology) or even dynamically 
(\emph{reconfigurable} topology) over time. 
We will refer to the latter also as  \textbf{\emph{self-adjusting networks}}.
While first empirical studies show 
that a demand-aware network can 
achieve performance similar to a demand-oblivious
network \emph{at lower cost}~\cite{firefly,projector},
not much is known today about the \emph{algorithmic problems}
underlying the design of self-adjusting networks. 
Indeed, while reconfigurable networks introduce an interesting
paradigm shift, we currently lack analytical tools to investigate their
potential and implications. 

\begin{figure}[t]
\centering
 \vspace{.5cm}
\includegraphics[width=.95\columnwidth]{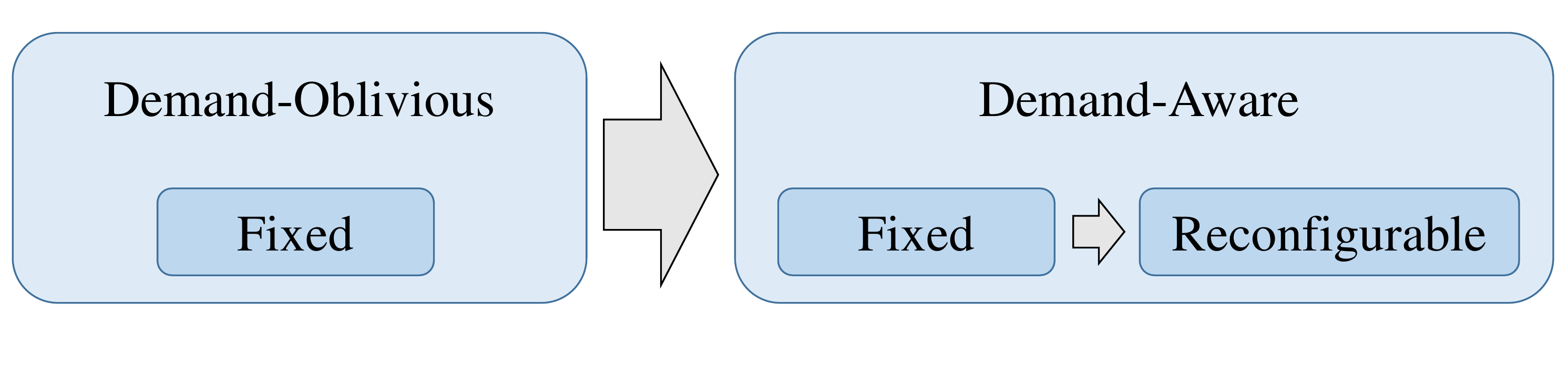} \\ 
\caption{Simple taxonomy of network optimization}
\label{fig:evolution}
\end{figure}

This paper initiates the study of the theory
of demand-aware, self-adjusting networks, and in particular
their fundamental underlying algorithmic problems.
Our position is that self-adjusting networks should be seen from the perspective
of self-adjusting datastructures: 
The current paradigm shift toward ``self-optimizing'' network topologies resembles
the process that data structures went through over 40 years ago~\cite{splaytrees},
evolving from static worst-case designs toward 
demand-aware and then self-adjusting designs, see Fig.~\ref{fig:evolution}.

\begin{figure*}[t]
\centering
\begin{tabular}{ccc}
\includegraphics[width=.65\columnwidth]{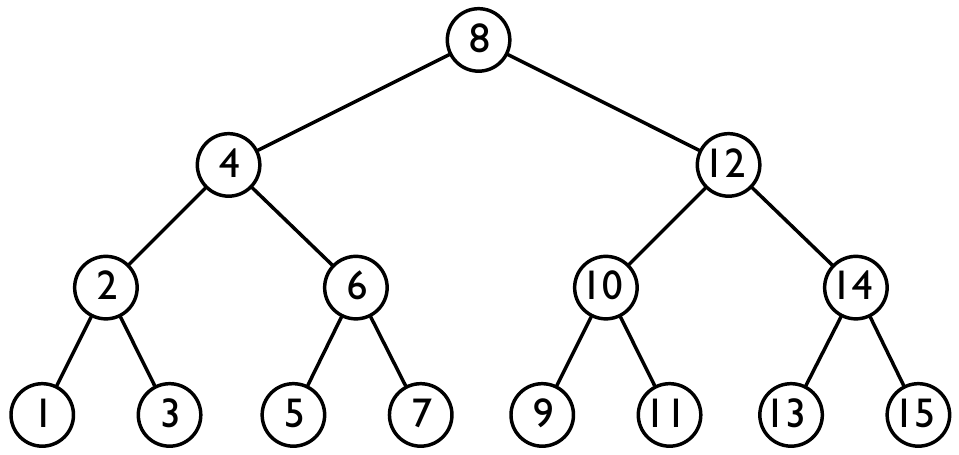} &
\includegraphics[width=.50\columnwidth]{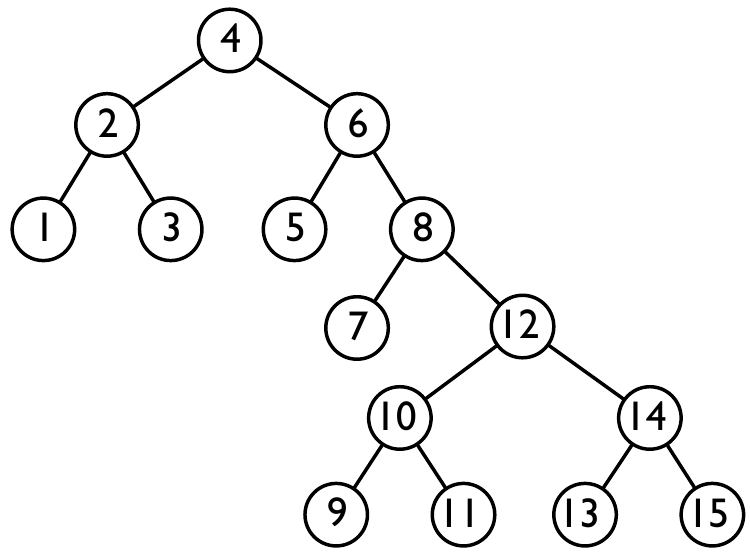} &
\includegraphics[width=.80\columnwidth]{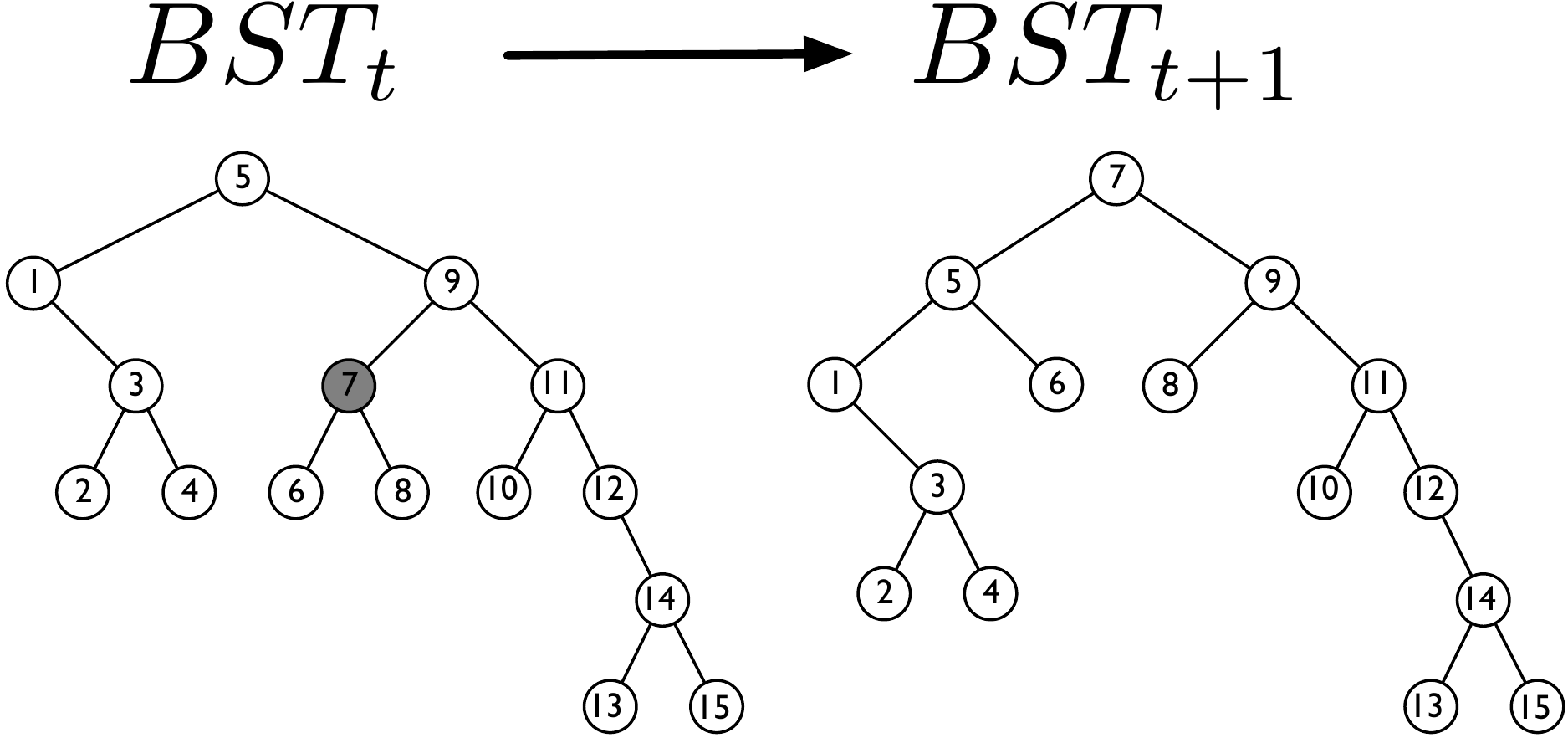}  \\
(a) BST: demand-oblivious & (b) BST: demand-aware & (c) BST: self-adjusting\\ \\
\end{tabular}
\caption{Examples of different demand graphs and their corresponding networks}
\label{fig:exampleBST}
\end{figure*}

As an illustrative and well-studied example, consider the 
case of Binary Search Trees (BSTs), see
Fig.~\ref{fig:exampleBST}. 
Traditional BSTs are \emph{(demand-)oblivious} and do not rely on
any assumptions on the demand (e.g., lookup requests) they serve, but optimize 
for the \emph{worst-case}, where \emph{any} item could be 
accessed: items are stored 
at distance $O(\log{n})$ from the root, uniformly and independently
of their frequency, see Fig.~\ref{fig:exampleBST}~(a).

Clearly, if the demand has a specific pattern, 
the performance of the binary search tree designed for the worst case, 
is no longer optimal. \emph{Demand-aware}~(statically-)optimized but
\emph{fixed} BSTs (a.k.a.~biased search trees)
such as~\cite{Mehlhorn75,knuth1971optimum,hu1971optimal,bent1985biased} 
account for the frequency 
of the accessed items: frequent items are stored close to the root, infrequent items are lower in the tree, see Fig.~\ref{fig:exampleBST}~(b).

\emph{Self-adjusting} BSTs, or dynamic demand-aware BSTs, 
are an attractive alternative to fixed BSTs, as they do not rely on an a priori 
knowledge about the demand.
Rather, self-adjusting BSTs learn and adjust to the demand, and to its
\emph{temporal locality}, in an 
\emph{online manner}. 
This, by now classical, approach was first introduced 
by Sleator and Tarjan~\cite{splaytrees}
for \emph{splay trees}, and today, several other self-adjusting BSTs 
exists, such as \emph{tango trees}~\cite{tangotrees}.
As we will discuss later, despite not knowing the demand ahead of time, 
self-adjusting BSTs ideally never perform much worse than any fixed tree, 
but can perform significantly better if the demand features spatial or temporal locality.

In the same spirit, we in this paper present a taxonomy and a formal model for \emph{Self-Adjusting Networks (SANs)}. 
We show that while the performance of demand-oblivious networks
is limited by worst-case metrics such as the network diameter,
self-adjusting networks are only limited by the 
spatial and temporal locality. The more ``structure'' the demand has,
the better self-adjusting networks can perform compared to
demand-oblivious networks. 
We demonstrate by examples 
the inherent benefit of demand-aware
networks over state-of-the-art demand-oblivious networks such as
expander graphs, identify objectives, define desirable properties and metrics
of demand-aware networks, and discuss open problems.



\begin{figure*}[t!]
\centering
\begin{tabular}{ccc}
\includegraphics[width=.45\columnwidth]{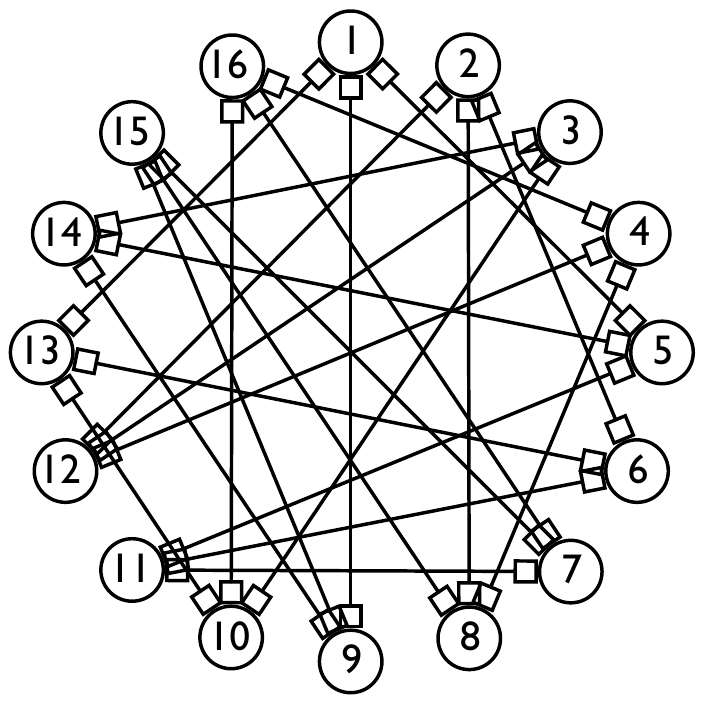} ~~~&~~~
\includegraphics[width=.35\columnwidth]{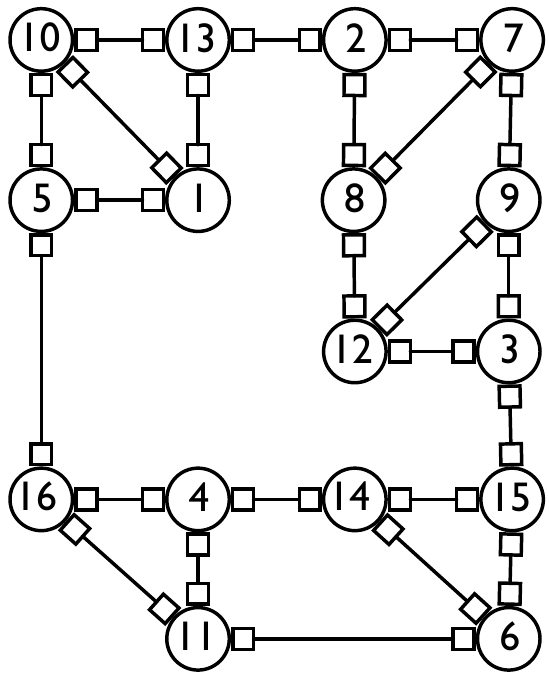} ~~~&~~~
\includegraphics[width=.80\columnwidth]{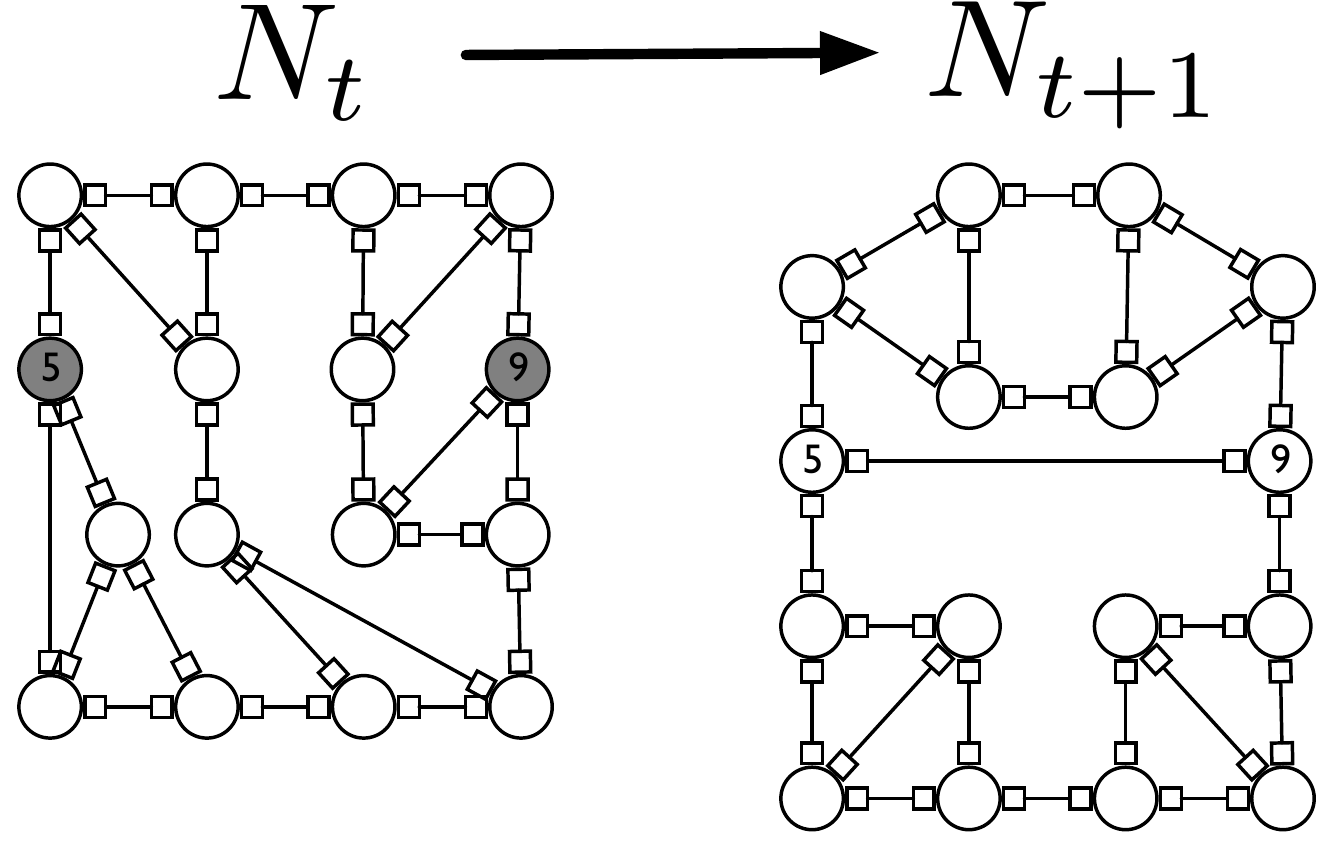}  \\
 (a) Network: demand-oblivious & (b) Network: demand-aware & (c) Network: self-adjusting
\end{tabular}
\caption{Examples of different demand graphs and their corresponding networks.}
\label{fig:exampleNet}
\end{figure*}

\section{Why Self-Adjusting Networks?}\label{sec:perspective}


The vision of self-adjusting networks can be best understood by an analogy
to datastructures. This section establishes and motivates this connection, 
and elaborates on the example of
Binary Search Trees (BSTs) and a case study of routing. 
We will first demonstrate the benefits 
of self-adjusting BSTs and later extend the example to self-adjusting networks.

\subsection{From Self-Adjusting Datastructures...}

We can identify three different kinds of binary search tree
datastructures: demand-oblivious BSTs, (static) demand-aware 
BSTs, and (dynamic) demand-aware BSTs, henceforth also called
self-adjusting BSTs.  We assume that the BST 
stores a set of items $\{1,2,\ldots,n\}$ where $n=2^{k}-1$.

Let us start by focusing on the most basic operation supported by a BST:
the $\search()$ operation (for simplicity, we ignore $\insrt()$ and $\delete()$).
The cost of a search for an item $v$ is proportional to the depth of
$v$ from the root of the BST (e.g., the number of 
pointer accesses). In graph terms, regarding the BST as a network,
this corresponds to the shortest path length
between the root and the searched element, $v$. 

Now consider a \emph{demand} to be served
by the BST, given as a \emph{sequence of $m$ search requests} 
$\tau=(\tau_0, \tau_1, \dots \tau_{m-1})$ for items (i.e., \emph{keys}).
$\tau$ is the problem \emph{input}. 
For demonstration purposes, in the following, we will assume the specific search sequence 
$\tau= (1, 1, \dots , 1, 3, 3, \dots , 3, 5, 5, \dots , 5, \ldots, \ldots, 2k-1, 2k-1, \dots, 2k-1)$ 
where an item is repeated many times consecutively. Note that $\tau$ contains $k=\log n$ unique items which correspond to the $k$ smallest leaves in the complete BST, see Fig.~\ref{fig:exampleBST}~(a).

Different BSTs, depending on whether they are demand-oblivious, static 
(\emph{fixed}) demand-aware, and dynamic (\emph{reconfigurable}) 
demand-aware, will incur different costs under this demand. 
In the following, we will examine the three different cases in turn.

\noindent \textbf{Demand-Oblivious Datastructures.}
Let us first study a demand-oblivious BST, 
namely a tree that is designed to perform well even for a ``worst possible''  $\tau$. 
A balanced and complete BST over the items $1, \dots n$
provides an optimal solution for the demand-oblivious case, see 
Fig.~\ref{fig:exampleBST}~(a).
Such a tree guarantees a cost of at most $\log n$ for every request 
in every sequence. Note that $\log n$ is also the \emph{maximum empirical 
entropy} of an $n$-items sequence, where all items have the same
(uniform) frequency.
For our specific sequence $\tau$, which is unknown a priori, 
the amoritzed cost per request will be $\log n$: 
all items are leafs in the complete tree. 

\noindent \textbf{Demand-Aware Datastructures.}
Next, consider a (fixed) demand-aware BST which is optimized toward 
$\tau$ a priori, 
like in Fig.~\ref{fig:exampleBST}~(b). Such a demand-aware, optimized tree, 
can take advantage of the \textbf{\emph{spatial locality}} of the demand, 
and will put all the $\log n$ requested items near the root 
(and other elements further away), resulting in an amoritzed 
cost of only about $\log \log n$ per request. 
Such optimized demand-aware trees have been studied, e.g., by 
Knuth~\cite{knuth1971optimum} and Hu et al.~\cite{hu1971optimal}
who presented polynomial-time algorithms to construct exactly optimal trees
for given probability distributions,
as well as by Mehlhorn~\cite{Mehlhorn75} and Bent et al.~\cite{bent1985biased} who
presented faster algorithms for approximately optimal trees. 
The amoritzed cost per request in these trees is proportional to the \emph{empirical entropy} of the sequence, $ \hat{H}(\tau)$. The empirical entropy is always
$ \hat{H}(\tau) \le \log n$, and it can be much lower than $\log n$: 
in our example, $\hat{H}(\tau) \approx \log k =  \log \log n$.

\noindent \textbf{Self-Adjusting Datastructures.}
To conclude the example, let us consider a self-adjusting BST, 
as shown in Fig.~\ref{fig:exampleBST}~(c). For this case,
the sequence is \emph{unknown} a priory, but we can self-adjust 
the tree between requests. For every time $t$, we consider a (possibly) 
different binary search tree $BST_t$: we need a new operation, $\adjust()$, 
which reconfigures $BST_t$ to $BST_{t+1}$. Such an adjustment obviously 
comes at a \emph{cost}, and is usually implemented using \emph{local tree rotations}
(each of constant cost) which preserve the search structure of the BST. 
More specifically, a tree rotation can only be performed for an accessed item and only by changing pointers with immediate neighbors (i.e., parents, children in the tree). Splay trees~\cite{splaytrees} for example, use a ``move-to-front'' rule, 
where the last requested item is rotated to the root, using tree rotations known as \emph{splay operations}.

For the above considered sequence $\tau$, the amortized cost per request (including both $\search()$ and $\adjust()$) will be \emph{constant}. Each requested item will move-to-front once, at high cost, but then this cost will be amoritzed by the subsequent repetitions of requests for the same item, taking advantage of  \textbf{\emph{temporal locality}}.
Surprising at first, but by now well-known, is that for \emph{any sequence}, 
splay trees are \emph{statically optimal}: they perform as well as any demand-aware tree that is a priori optimized toward the demand, like Mehlhorn trees.

To summarize the BST example, for the above toy sequence~$\tau$, the amortized cost per request will be about $\log n$ (for oblivious BSTs), $\log \log n$ (for demand-aware but static BSTs), or even \emph{constant}~(for self-adjusting BSTs), depending on the kind of BST. This clearly demonstrates the possible cost benefits of demand-aware and self-adjusting datastructures. For example, self-adjusting BSTs are useful for implementing caches and garbage collection where the \emph{principle of locality}~\cite{denning2005locality} holds.

\subsection{... to Self-Adjusting Networks}

We now repeat the same motivation for networks. 
We look at a network as a datastructure that,
rather than serving search requests 
issued from the root to an item, 
serves \emph{communication 
requests}~(e.g., packets) from a \emph{source} node to a \emph{destination} node. 
This operation is performed abstractly, via a $\route()$ operation, 
similar in spirit to the BST's $\search()$ operation (where the source
is always the root).
The input to this routing problem is a sequence $\sigma=(\sigma_0, \sigma_1, \dots, \sigma_{m-1})$ of communication requests, where each request amounts to forwarding one unit of data from a source node $v$ to a destination node $u$.
While network optimization in general obviously has many dimensions, and includes aspects such as addressing, policies, congestion,  etc., in the following, we will only consider the \emph{routing} or \emph{forwarding} cost of each request: the cost of serving a request $\sigma_i$ is given by the \emph{length} of the route
from source to destination. In particular, 
in our example, we will assume \emph{shortest path} routing. 

\noindent \textbf{Demand-Oblivious Networks.} We start with the topologies of traditional \emph{demand-oblivious} communication networks,
which do not rely on any assumptions on the demand,~$\sigma$.
Rather, they are conservatively optimized for \emph{arbitrary}~(i.e., all-to-all)
demands, providing worst-case properties
such as bounded network diameter, mincut, 
or (almost) full bisection bandwidth (even in the presence of traffic engineering flexibilities~\cite{fat-free}).
Fig.~\ref{fig:exampleNet}~(a) presents an example for such a state-of-the-art network, an \emph{expander}-based network~\cite{hoory,xpander}.

\begin{figure}[t]
\centering
\begin{tabular}{cc}
\includegraphics[width=.45\columnwidth]{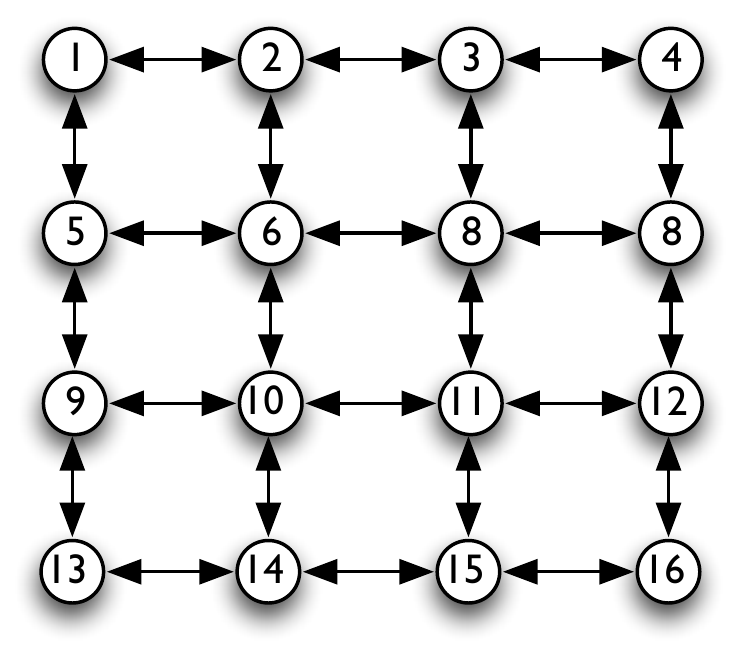} &
\includegraphics[width=.42\columnwidth]{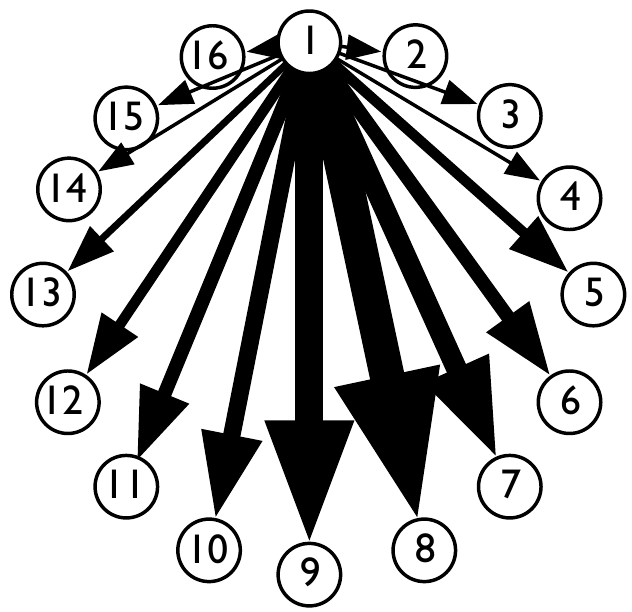} \\
(a) & (b) 
\end{tabular}
\caption{Expander networks do not achieve optimal average 
route lengths for sparse demand graphs.
(a)~An oblivious embedding of a 2-dimensional grid demand graph on a constant degree expander network  will result in average
route lengths of
$\Omega(\log n)$, while the conditional entropy of the demand graph 
is less than two.
(b) An oblivious embedding of a weighted star demand graph
on a constant degree expander network will result in 
an average route length of~$\Omega(\log n)$, 
while the conditional entropy of the demand graph could be much lower.
}
\label{fig:example3}
\end{figure}

What will be the amortized cost (i.e., the average route length) 
per request on such an expander?
To start with a simple example (and for the sake of simplicity and 
clarity, we leave out some of the details),
consider a demand $\sigma$ whose communication
pattern is described by a two-dimensional square grid, of size
$\sqrt{n}\times \sqrt{n}$ (see Fig.~\ref{fig:example3}~(a)). We call 
this representation a \emph{demand graph}~$\demg(\sigma)$ where each 
weighted (directed) edge $e=(v, u)$ in the graph represents 
the frequency at which the two endpoints of $e$, namely $v$ and $u$, 
communicate in $\sigma$.
Note that in this request sequence $\sigma$, 
every node communicates with at most four partners, hence, 
it is a \emph{sparse} sequence, with \emph{spatial locality}.

Serving this demand 
on a static expander 
in an oblivious (i.e., arbitrary) way
will result in an 
average route length in the order of~$\log n$, 
the diameter of a bounded degree expander.
Note again that  $\log n$ is the \emph{maximum empirical  entropy} 
of the demand, $H(\sigma)$.

\noindent \textbf{Demand-Aware Networks.}
What about \emph{demand-aware} networks? Can the average route length be better than $\log n$ if the network is optimized toward the demand $\sigma$ \emph{known a priori}, as in Fig.~\ref{fig:exampleNet}~(b)? It turns out that the answer is affirmative for many cases, as was shown recently in~\cite{dan}. A fundamental metric for the performance of such demand-aware networks turned out to be the (empirical) \emph{conditional entropy} of $\sigma$. In a nutshell, the conditional entropy is a measure of the \emph{spatial locality} of $\sigma$. In our example in  Fig.~\ref{fig:example3}~(a),
since every node 
communicates with at most four partners (other nodes),
the conditional entropy is 
\emph{a constant}.
In other words, there is a large gap of~$\Theta(\log{n})$ between the 
conditional entropy and the entropy of $\sigma$.
Clearly, a demand-aware network can be designed to serve our $\sigma$ at a very low 
(amortized) cost per request. The results in~\cite{dan} prove that if 
$\demg(\sigma)$ is sparse, then the conditional entropy is both a lower and an upper bound for the average route length; the paper presents a design that matches 
the upper bound.
Another example introducing a large gap of~$\Theta(\log{n})$
between demand-oblivious and demand-aware networks 
is a demand graph~$\demg(\sigma)$ which forms a star 
(with \emph{unbounded} degree), see Fig.~\ref{fig:example3}~(b): 
node pairs communicate at different frequencies (skewed distribution,
as indicated by the thickness). For this demand, the conditional entropy could be much lower than
$\log n$ which will be the cost of serving this demand on an oblivious expander.

More generally, one can see that 
every sparse communication pattern
which is
embedded on a demand-oblivious expander,
will result in average route lengths in the order of~$\Omega(\log{n})$, 
the diameter, regardless of the entropy or the conditional entropy of the demand. 

\noindent \textbf{Self-Adjusting Networks.}
We complete the analogy by moving to self-adjusting networks, see  Fig.~\ref{fig:exampleNet}~(c).
Similar to the BST case, we have a new operation, $\adjust()$, to reconfigure the network at time $t$, $\netw_t$, to a new network at time $t+1$, $\netw_{t+1}$. This reconfiguration will also come at a cost that needs to be well-defined (and to be compared to the routing cost).

Like in BSTs,  we can ask: is there a design of a \emph{self-adjusting} network that achieves the bounds 
of an optimal  fixed network,
\emph{without knowledge of~$\sigma$},
but using reconfigurations in an \emph{online} manner? 
In other words, are there \emph{statically optimal} self-adjusting networks? Like splay trees are for binary search trees?

Moreover, similarly to BSTs, we note that  
self-adjusting networks, taking advantage both of \emph{spatial} and \emph{temporal} locality, can in principle perform much better than
existing cost lower bounds (such as~\cite{dan}) for \emph{static} demand-aware networks. For example one can think of requests that arrive from a grid like in Fig.~\ref{fig:example3}~(a), but where the grid also changes over time, to add temporal locality. For this case, self-adjusting networks will perform better than static networks.

\section{Taxonomy}\label{sec:taxonomy}

\begin{figure*}[t]
\centering
\includegraphics[width=.7\textwidth]{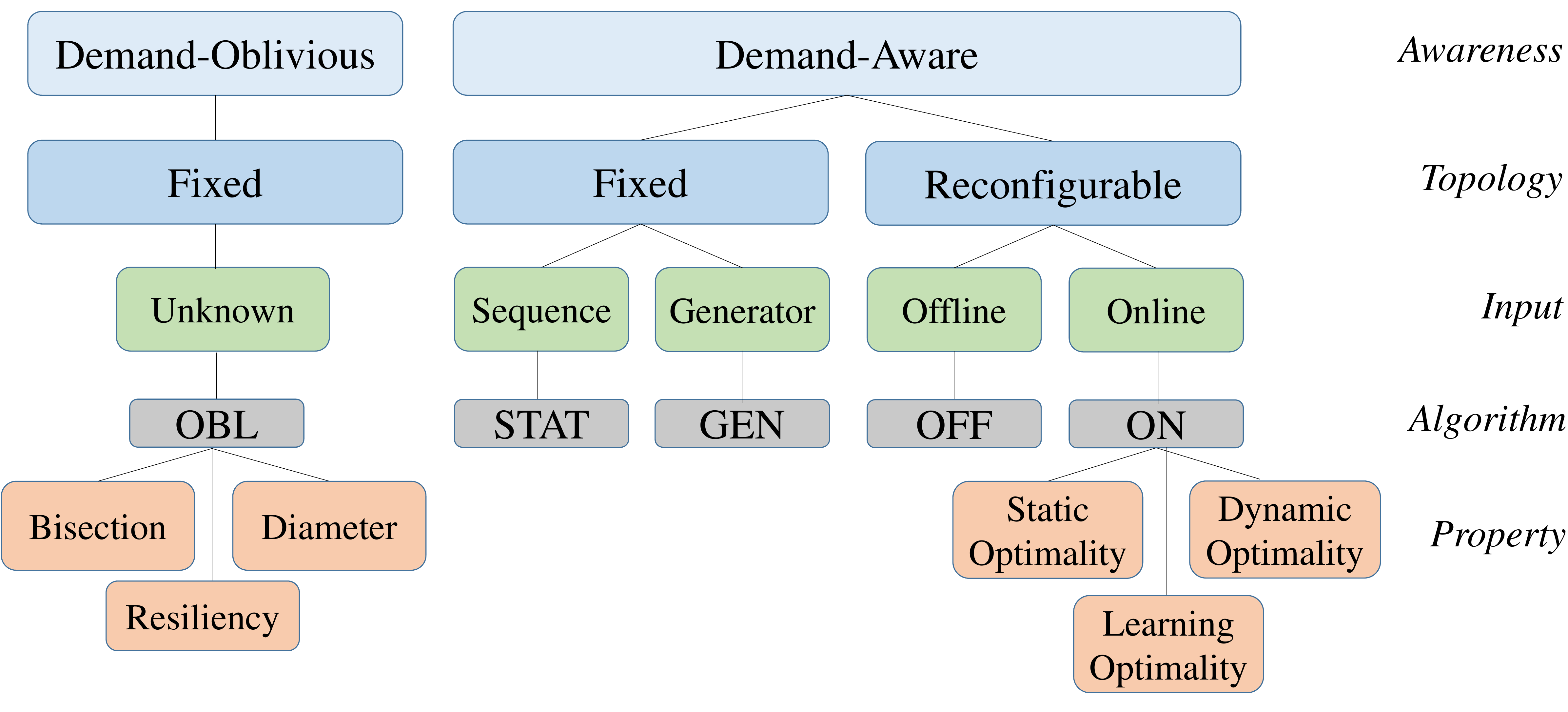} \\ 
\caption{Detailed taxonomy of network optimization}
\label{fig:taxonomy}
\end{figure*}

This section presents a more systematic taxonomy
of network designs, revolving around the \emph{(demand) awareness},
the type of \emph{topology}~(fixed or reconfigurable), the
type of \emph{input}~(e.g., unknown, known, revealed online over time),
as well as the required \emph{algorithms} and \emph{properties}.
See Fig.~\ref{fig:taxonomy}.

\subsection{Demand-Oblivious Networks}

State-of-the-art demand-oblivious datacenter networks such as Xpander~\cite{xpander} rely on a \emph{fixed}~(static)
network topology and are not optimized toward a specific demand: 
the demand (i.e., input) is \emph{unknown}.
Typical objectives of such network designs are to provide (almost)
full bisection bandwidth (assuming all-to-all communication patterns), 
short routes (e.g., at most 6 hops from server to top-of-the-rack
switch, aggregation switch, core, and down again), as well as resiliency
(e.g., $k$-connectivity).
We will refer to algorithms for demand-oblivious networks by
$\Obl$.

\subsection{Demand-Aware Networks}

\noindent \textbf{Fixed Demand-Aware Networks.}
The input to the fixed (static) demand-aware network design
problem could either be a sequence of requests 
(in our case, communication requests) $\sigma$
or a generative model $\mathcal{G}$ (\emph{generator}) of such requests.
A generator $\mathcal{G}$ comes with a set of parameters $\params(\mathcal{G})$.
A most simple example for a generative model is a \emph{fixed
distribution} 
from which requests are sampled \emph{i.i.d.}:
such a generator may feature spatial locality, however, it does not feature
any temporal locality as requests are sampled independently. 
A more complex generative model which also features temporal
locality could be, for example, a \emph{Markovian process}~(or random walk). 

Independently of whether the input is a sequence or
generator, our goal is to design an optimal \emph{fixed} topology 
$N^*$, and we will refer to the corresponding algorithms as 
$\Fixed$ and $\Gen$.


\noindent \textbf{Reconfigurable Demand-Aware Networks.}
Reconfigurable demand-aware networks allow to optimize
the topology at runtime, i.e., the topology is dynamic. 
If the demand is known a priori, an offline algorithm $\Off$ can be used to compute
an optimized schedule to reconfigure a network over time, i.e., 
unlike $\Fixed$, $\Off$ changes the network over time and is charged
for such reconfigurations. In other words, $\Off$ should only change the network
when it pays off later, by making requests cheaper to serve.

Most interesting are scenarios where the input
is not known a priori but revealed over time, in an \emph{online} 
manner. 
We distinguish between three important
metrics: \emph{static optimality}, \emph{learning optimality},
and \emph{dynamic optimality}.
We have already discussed static optimality above: it asks
for an online algorithm $\On$ which is competitive compared to an optimal static
algorithm $\Fixed$. In contrast, dynamic optimality asks for an algorithm $\On$ 
which is competitive even when compared to an optimal (dynamic) offline algorithm $\Off$, a much stronger requirement. 

An interesting special case regards a scenario in which the demand
is fixed but not known ahead of time, and
the goal of the online network reconfiguration algorithm  
is to \emph{learn} the demand generator quickly and cost-effieciently
(with little reconfigurations). 
For example, if the demand is given by a generator describing
a fixed distribution from which requests are sampled i.i.d.,
ideally, an algorithm would quickly converge
to an optimized fixed network which optimally serves this
distribution: since requests are i.i.d., there is no temporal locality
and reconfigurations are no longer beneficial after convergence.  
If the generator is a Markov process, the request sequence may feature
both spatial and temporal locality, and an algorithm should quickly learn
the process and may then converge to an optimal reconfiguration schedule
over time. 
We hence introduce a third type of optimality besides
static and dynamic optimality:  
\emph{learning optimality}. Learning optimality asks
for an online algorithm $\On$ which is competitive compared to an optimal 
static algorithm $\Gen$ which knows the generator. 

Finally, we can classify algorithms for online reconfigurable demand-aware networks
in two flavors: centralized algorithms and
distributed (i.e., decentralized and concurrent) algorithms.
 
\subsection{Additional Properties}

Besides the properties that are specific to
demand-aware networks, it is usually desirable that 
demand-aware networks additionally still fulfill the 
 traditional properties of demand-oblivious networks,
 for example the requirement to provide redundant
connectivity. 
Furthermore, some static properties become
more useful in the dynamic conext, for example, \emph{compact and local
routing}: 
As dynamic demand-aware networks may change frequently over time,
it may be highly undesirable to recompute routing paths each time
for each topological modification; rather, it would be ideal
if the topology allows to forward packets greedily, at any time,
and modifications only entail local changes to the forwarding tables.

\section{A Formal Model}\label{sec:model}

This section presents a general algorithmic model 
for self-adjusting networks. 
We consider a set of~$n$ nodes 
$V=\{1,\ldots,n\}$
(e.g., the top-of-rack switches). 
The communication \emph{demand} among these nodes
is a sequence 
$\sigma =
(\sigma_1, \sigma_2, \ldots)$ of \emph{communication
requests} where~$\sigma_t = (u,v)  \in V \times V$, 
is a source-destination pair. 
The communication demand can either
be finite or infinite.

In order to serve this demand, the nodes~$V$ must
be inter-connected by a network~$\netw$, defined over the 
same set of nodes. In case of a demand-aware network,
$\netw$ can be optimized towards~$\sigma$,
either statically or dynamically:
a self-adjusting network~$\netw$ can change over time, and we denote
  by~$\netw_t$ the network at time~$t$, i.e.,
	the network evolves: $\netw_0, \netw_1, \netw_2, \ldots$ 
	
\subsection{Constraints}

In addition to the dynamic properties related to optimizations over time,
described shortly, 
a network $\netw_t$ may have to adhere to some physical constraints 
(e.g., the number of lasers which can be installed on a top-of-the-rack
switch may be limited) and 
fulfill invariants at any time.
This can be modeled by requiring
that all networks $\netw_{t}$ belong to some network family 
$\netws$: $\netw_{t}\in \netws$. 
Examples for families $\netws$ may include, 
\emph{bounded degree} networks (e.g., for a high scalability), networks of 
\emph{full bisection bandwidth} or  \emph{expanders}
(e.g., to ensure congestion-free
shuffle phases),
\emph{$k$-connected} networks (for resiliency), etc.

\subsection{Reconfiguration}

The crux of designing smart self-adjusting networks is 
to find an optimal \emph{tradeoff} between the benefits
and the costs of reconfiguration:
while by reconfiguring the network, we may be able to serve
requests more efficiently in the future, reconfiguration
itself can come at a cost.

The inputs to the self-adjusting network design problem 
is a set of allowed network topologies $\netws$, 
the request sequence $\sigma=(\sigma_0,\sigma_1,\ldots,
\sigma_{m-1})$, and two types of costs:
\begin{itemize}
\item An \textbf{adjustment cost} $\RecCost: \netws \times \netws \rightarrow \mathbb{R}$ which defines the cost of reconfiguring a network
 $\netw$ to a network  $\netw'$. Adjustment  costs may include
 mechanical costs (e.g., energy required to move lasers or abrasion) as well as 
 performance costs  (e.g., reconfiguring a network may entail control plane
 overheads or packet reorderings,
 which can harm throughput).  
 For example, the cost could be given 
by the number
of links which need to be changed in order to transform
the network. 
\item A \textbf{service cost} $\RouCost: \sigma \times
\netws \rightarrow \mathbb{R}$ which defines,
for each request $\sigma_i$ and for each network $\netw\in \netws$,
what is the price of serving $\sigma_i$ in network $\netw$.
For example, the cost could correspond to the route length:
shorter routes require less resources and hence reduce not only
load (e.g., bandwidth consumed along fewer links),
but also energy consumption, delay, and flow completion
times, could be considered for example.
\end{itemize}

Serving request $\sigma_i$ under the current
network configuration $\netw_{i}$
will hence cost $\RouCost(\sigma_i,\netw_i)$,
after which the network reconfiguration algorithm may decide
to reconfigure the network at cost $\RecCost(\netw_{i},\netw_{i+1})$. 
The total processing cost of a schedule $\sigma$
is then
$$
\Cost(\A, \netw_0, \sigma) = \sum_{i=0}^{m-1} \RouCost(\sigma_i,\netw_{i}) 
+ \RecCost(\netw_{i},\netw_{i+1}) 
$$
\noindent where $\netw_i\in \mathcal{N}$ denotes the network at time $i$.


It is sometimes useful to aggregate the requests of sequence 
$\sigma$ over time and represent it 
as a directed and weighted \emph{demand graph}~(resp.~guest graph or request graph) $G(\sigma)=(V(\sigma),E(\sigma))$. 


We need the concept of amortized costs to reason about
costs over sequences.  
\begin{definition}[\bf Average and Amortized Cost]
Given an algorithm~$\A$, an initial network~$\netw_0$, a 
reconfiguration cost function $\RecCost$, a request serving cost
function $\RouCost$, and a sequence~$\sigma=(\sigma_0,
\sigma_1, \ldots, \sigma_{m-1})$ of communication requests over time,
we define the \emph{(average) cost} incurred by~$\A$ as:
$$
\Cost(\A, \netw_0, \sigma) = \frac{1}{m} \sum_{i=0}^{m-1} 
 \RouCost(\sigma_i,\netw_{i})
 +
 \RecCost(\netw_{i},\netw_{i+1}) 
$$
\noindent where $\netw_i\in \mathcal{N}$ denotes the network at time $i$.
The \emph{amortized
cost} of~$\A$ is defined as the worst possible cost of~$\A$
 over all initial networks~$\netw_0$ and all 
 sequences~$\sigma$, i.e.,
$\max_{\netw_0,\sigma}\Cost( \A, \netw_0, \sigma)$.
\end{definition}


\subsection{Objectives and Metrics}\label{sec:routing}

We can now define static, dynamic, and learning optimality objectives. 
In \emph{static optimality}, want the network to (asymptotically) perform 
well \emph{even in hindsight}, i.e., \emph{given}
knowledge of the demand. 

\begin{definition}[\bf Static Optimality]
Let~$\Fixed$ be an optimal static algorithm with perfect
knowledge of the demand~$\sigma$,
and let~$\On$ be an online algorithm. We say that~$\On$ is \emph{statically
optimal}
if, for sufficiently long communication patterns~$\sigma$, the following
ratio is \emph{constant}:
$$
\rho = \max_{\sigma} 
\frac{\Cost(\On, \netw_0, \sigma)}{\Cost(\Fixed,\netw^*, \sigma)} +\beta
$$
\noindent
for some $\beta$ independent of the length of the sequence $\sigma$.
Here,~$\netw_0 \in \mathcal{N}$ is the initial network, 
from which ~$\On$ starts,
and~$\netw^* \in \mathcal{N}$ is the statically optimal network. 
In other words,~$\On$'s cost is at most a 
constant factor higher than~$\Fixed$'s in the worst case.
\end{definition}

The holy grail of self-adjusting networks however regards 
the design of \emph{dynamically} optimal
reconfigurable networks: how well can a reconfigurable
demand-aware network perform, when compared to a network
which is dynamically optimized in an offline manner?

\begin{definition}[\bf Dynamic Optimality]
An algorithm is called \emph{dynamically optimal} if and only if it 
is asymptotically optimal even compared to an optimal
offline algorithm which can dynamically reconfigure the network
and which has complete knowledge of the request sequence~$\sigma$
\emph{ahead of time}.
More formally, let~$\Off$ be an optimal offline algorithm,
and let~$\On$ be an online algorithm. We say that~$\On$ is \emph{dynamically
optimal}
if the following ratio is \emph{constant}:
$$
\rho = \max_{\sigma} \frac{\Cost(\On, \netw_0, \sigma)}{\Cost(\Off, \netw_0, \sigma)}
$$
\noindent 
that is,~$\On$'s cost is at most~$\rho$ times higher in the worst case.
Again, $\netw_0 \in \mathcal{N}$ is the initial network.
\end{definition}

Finally, we define learning optimality which lies between the above:
\begin{definition}[\bf Learning Optimality]
An algorithm (which initially does not know the parameters $\params(\mathcal{G})$
of the generator) is called \emph{learning optimal} if and only if it 
is asymptotically optimal even when compared to an optimal
static algorithm which
\emph{knows the generator}.
More formally, let~$\Fixed$ be an optimal fixed algorithm, 
and let~$\Gen$ be an online learning algorithm. 
We say that~$\Gen$ is \emph{learning
optimal}
if the following ratio of expectations is \emph{constant}:
$$
\rho = \max_{\params(\mathcal{G})} \frac{\mathbb{E}[\Cost(\On, \netw_0, \sigma)]}{\mathbb{E}[\Cost(\Gen, \netw^*, \sigma)]} +\beta
$$
\noindent where $\beta$ is independent of the length of the sequence
and where the maximum is taken over the parameters
of the generator model $\mathcal{G}$.
That is,~$\Gen$'s cost is at most~$\rho$ times higher in the worst case.
\end{definition}

%




\section{Review of State-of-the-Art}\label{sec:relwork}

The problem of designing demand-aware and self-adjusting networks is a fundamental one, 
and finds interesting applications in many distributed and networked systems,
not only in datacenters.
For example, use cases also arise
in the context of wide-area networks~\cite{Jia2017,jin2016optimizing}
and, more traditionally, in the context of overlays~\cite{scheideler2009distributed,ratnasamy2002topologically}.
However, while the problem is natural, surprisingly little is known today about the design of
demand-aware networks, especially dynamic networks which can change over time.
The approach of reconfiguring
network topologies to reduce communication
costs, is orthogonal to approaches changing the
traffic matrix itself (e.g.,~\cite{fibium}) or 
migrating communication endpoints on a fixed topology~\cite{disc16}.

One basic observation to make is that the design of static demand-aware networks
is related to graph embedding problems (a.k.a.~virtual network
embedding and graph layout problems)~\cite{diaz2002survey,ifip18round,bansal2011minimum}: given a graph (describing
the demand), find an embedding in another graph
(the physical topology), such that certain properties are fulfilled
(e.g., the sum or max of the total load on the physical graph
is minimized). It is known that the embedding problem is
NP-hard in many variants~\cite{ifip18landscape}, for example
already for very simple physical networks such as the line~\cite{diaz2002survey}:
the problem of embedding an arbitrary request graph onto a line
is known as the Minimum Linear Arrangement (MLA) problem.
From this relationship it also follows that the static offline  
demand-aware network design problem is NP-hard if the designed
network is restricted to the family of degree-2 networks. 

However, unlike graph embedding problems, in the design of
demand-aware networks, the physical network is not given but
subject to optimization as well. An intriguing and open question is whether this
additional degree of freedom makes the problem harder or easier.
An encouraging example (focusing on routing) is given in~\cite{splaynets}:
an optimal fixed demand-aware network (also called \emph{DAN}
in the literature) restricted to BSTs
can be computed in polynomial time. 
Another example of approximately optimal demand-aware networks
are $\textsc{Dan}$s~\cite{dan}~(which provide a constant approximation
for sparse demand graphs), matching the lower bounds based on 
conditional entropies derived in the same paper.
A slightly different model, motivated by optical switches which can be 
configured to provide an optimized matching, is studied in~\cite{ancs18}:
the authors present several optimal algorithms for special workloads (e.g., where the demand is given by a single flow).
There is also work on resilient demand-aware networks, such as $\textsc{rDan}$~\cite{rdan}~(based on a coding approach but
without degree bounds).

Even less is known about \emph{self-adjusting} networks. 
The best upper bound known so far  for online reconfigurable
networks
is~$O(\hat{H}(X_{\sigma}) + \hat{H}(Y_{\sigma}))$, 
where~$\hat{H}(X_{\sigma})$ and~$\hat{H}(Y_{\sigma})$ are the
empirical entropies of sources and 
destinations in~$\sigma$, respectively~\cite{ton15splay}. 
It is achieved by a self-adjusting \emph{tree} 
network (the tree network can also be decentralized~\cite{displaynets}).
While this is optimal for some frequency distributions (e.g., 
the empirical distribution of $\sigma$), and in particular product frequency
distributions, 
 in general, 
it can be far from optimal.

\section{Conclusion}\label{sec:conclusion}


We hope that our paper can nourish the ongoing 
discussions on the benefits and limitations of such reconfigurable
network topologies, and we believe that our work opens
many interesting questions for future research.
On the algorithmic front, a first important open question 
 concerns the design of a 
\emph{self-adjusting} network that achieves the bounds of a static network,
\emph{without knowledge of~$\sigma$},
but using reconfigurations in an \emph{online} manner: 
are there \emph{statically optimal} self-adjusting networks,
like splay trees are for binary search trees?
Similarly, the design of \emph{learning optimal} and 
\emph{dynamically optimal} demand-aware networks 
remains an open problem. The latter is particularly challenging:
in the context of datastructures, the problem of designing dynamically
optimal BSTs has been an open problem for many decades already~\cite{tangotrees}.

On the modeling front, further refinements are required to account
for the specific costs incurred by a self-adjusting network. In particular, 
today, we lack good models for the cost of reconfiguration of networks: 
costs may not only accrue
in terms of, e.g., energy needed for the reconfiguration but also in terms of
performance: as routing over a continuously changing topology can be challenging,
it is desirable that (multi-hop) routing can be performed locally,
i.e., greedily, without the need for (distributed) forwarding table recomputations. 
This property is called \emph{local routing}. Furthermore, models need
to be extended to account for other quantitative aspects such as load.


%

\noindent \textbf{Acknowledgments.}
We thank Monia Ghobadi and Robert Tarjan
for inputs and discussions.

 
{
  \bibliographystyle{ieeetr} 
\bibliography{literature}
}

\end{document}